\documentclass[aps,twocolumn,groupedaddress,amsmath,amssymb,showpacs]{revtex4}
\usepackage{graphicx}

\usepackage{ae}
\usepackage{color}
\usepackage{bm}
\usepackage{float}

\newcommand{\be}{\begin{equation}} \newcommand{\ee}{\end{equation}}
\newcommand{\ord}{\mathcal{O}}
\newcommand{\lang}{\left\langle} \newcommand{\rang}{\right\rangle}

\begin{document}
\title{Determination of Structure Tilting in Magnetized Plasmas - Time Delay Estimation in Two Dimensions}

\author{D\'avid Guszejnov}
\email[]{guszejnov@reak.bme.hu}
\affiliation{Department of Nuclear Techniques, Budapest University of Technology and Economics, Association EURATOM, M\H{u}egyetem rkp. 9., H-1111 Budapest, Hungary}

\author{Attila Bencze}
\author{S\'andor Zoletnik}
\affiliation{MTA Wigner RCP, EURATOM Association, PO Box 49, H-1525 Budapest, Hungary}

\author{Andreas Kr\"amer-Flecken}
\affiliation{Institute of Energy and Climate Research - Plasma Physics, Forschungszentrum Jülich, Association EURATOM-FZJ, D-52425 Jülich, Germany}

\date{\today}

\begin{abstract}
Time delay estimation (TDE) is a well-known technique to investigate  poloidal flows in fusion
 plasmas. The present work is an extension of the earlier works of A. Bencze and S. Zoletnik 2005 and B. T\'al et al. 2011. From the prospective of the comparison of theory and experiment it seem to be important to estimate the statistical properties of the TDE based on solid mathematical groundings. This paper provides analytic derivation of the variance of the TDE using a two-dimensional model for coherent turbulent structures in the plasma edge and also gives an explicit method for determination of the tilt angle of structures. As a demonstration this method is then applied to the results of a quasi-2D Beam Emission Spectroscopy (BES) measurement performed at the TEXTOR tokamak.
\end{abstract}

\pacs{52.70.Ds}

\maketitle

\section{Introduction}

Turbulence plays a key role in the transport of energy and particles in hot magnetized plasmas \cite{Doyle}, but it is still not completely understood, despite intensive scientific investigation. Numerical simulations have shown that sheared flows play have a significant role in the controlling plasma turbulence \cite{Terry}, while one of the most significant experimental results of the last couple of years is the discovery of quasi-stationary \cite{Fujisawa, Bencze_Berta} and oscillating flows (zonal flows) \cite{Kramer-Flecken}. 

It is believed that the tilting of eddies could have a significant impact on the excitation of sheared flows \cite{Hidalgo}. Momentum transfer from turbulent structures to the main flow can be described by a negative eddy viscosity \cite{Starr}. One of the requirements for this negative viscosity behavior is the presence of some kind of irregularities in the spatial distribution of turbulent eddies such as non-circular shape and tilt. Structures are inherently tilted in the radial-poloidal plane since their emergence ($\alpha_B$ -- \emph{balooning angle}) and are further tilted by the sheared flows, resulting in a time dependent tilt angle ($\alpha$) \cite{Fedorczak_balloon}. Theoretical studies of the ITG modes in toroidal geometry highlighted that this ballooning angle determines the linear growth rate of the instability as $\gamma \propto \cos{\alpha_B}$ \cite{Kishimoto}, showing that the strongest modes are less tilted. Therefore the accurate measurement of the ballooning angle can give insight in the mode dynamics of the underlying instability.

The main goal of the present work is to give a well grounded time delay estimate (TDE) based method for the experimental estimation of the time evolution of coherent structure parameters, including the tilt angle in case of moderately sheared flows, where the structure parameters can be considered constant between observation points. In this discussion the nonlinear interaction between coherent structures is neglected, despite the fact that edge plasma interactions are mainly nonlinear, as correlation and TDE techniques rely on the assumption that events are independent. Thus it can recover the linear and quasi-linear behavior of the plasma. Our discussion includes the mathematical derivation of the expected TDE and its variance in two dimensions as well as the standard deviation of the tilt angle. The results are applicable for the calculation of the coherent structure parameters and flow modulations together with their errors, thus determining the significance of changes. 

The outline of the paper is as follows. In Sec. \ref{sec:model} the mathematical model will be described, along with its statistical properties. The analytical results are then compared against simulations in Sec. \ref{sec:simulation}. Finally, in Sec. \ref{sec:application} the model will be applied to quasi-2D BES data from the TEXTOR tokamak as a demonstration.

\section{Mathematical model}\label{sec:model}

Our goal is to give a heuristic description of coherent density structures in the edge plasma. For this we will assume a dominant scale on which coherent structures emerge -- in accordance with the experiments, which filter out small scale ($< 1\,\rm{cm}$) and short-living fluctuations -- and these structures take part in no significant nonlinear interaction during the timescale of the measurement ($\ord\left( 5\,\rm{\mu s}\right)$).

\begin{figure}[h]
\begin {center}
\includegraphics[width=\linewidth]{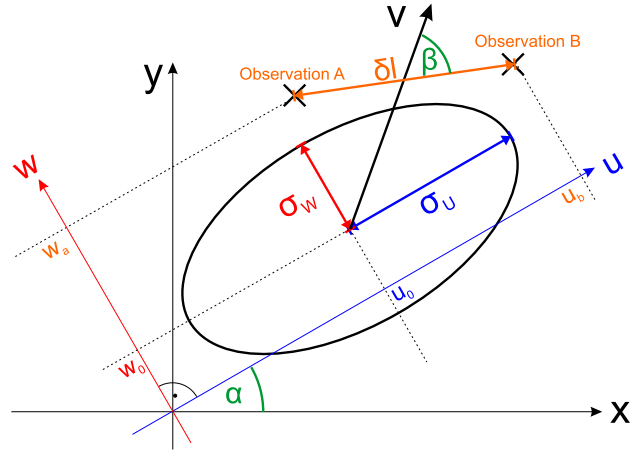}
\caption{Coordinate system used for the modeling of coherent structures, including the observation points.}\label{fig:model}
\end {center}
\end{figure}

For our analytic calculations we adopted a simple model, which assumes that the fluctuation of the plasma density is composed of small coherent structures. These have both Gaussian spatial distribution (in the direction of both their axes) and a Gaussian time decay as experiments have shown, that edge and core coherent structures exhibit Gaussian-like shape \cite{GarciaPPCF} (unlike SOL structures which can be highly asymmetric). The model also assumes that the coherent structures move at a constant velocity and have the same size and orientation. These assumptions are generally true for neighboring observation channels of turbulence measurements -- as the distance between them is usually 1-2 cms -- except for the cases of strongly sheared flows. This means that the density fluctuation caused by structure $i$ ($n_i$) can be expressed as
\begin{eqnarray}
\label{blob_eq}
n_i(u,w,t)=G(u,u_{i}+v_u (t-t_{i}),\sigma_U)\times\\
 G(w,w_{i}+v_w (t-t_{i}),\sigma_W) \times G(t,t_{i},\sigma_T)\nonumber,
\end{eqnarray}
where $u$, $w$ are coordinates in the coordinate system defined by the its axes (Fig. \ref{fig:model}), $v_u$, $v_w$ are the projected velocity components in these directions, while $G(x,x_{i},\sigma_x)$ denotes a Gaussian function defined as
\be
\label{gaussian}
G(x,x_0,\sigma)=\frac{1}{\sqrt{2 \pi} \sigma}e^{\frac{-\left(x-x_0\right)^2}{2 \sigma^{2}}}
\ee

If $N$ structures are present in the vicinity of two observation points ($[u_a;w_a]$ and $[u_b;w_b]$), then -- assuming linearity -- the local density can be written as
\be
\label{n_total}
n(u,w,t)=\sum_{i=1}^{N}{n_i(u,w,t)}.
\ee
From Eq. (\ref{n_total}) the cross correlation between the signals measured at point A and B can be formally expressed as
\begin{widetext}
\be
\label{cross_correlation_tot}
C(u_a,w_a,u_b,w_b,t)= \overline{(n(u_a,w_a,t)- \overline{n(u_a,w_a,t)}) (n(u_b,w_b,t+\tau)- \overline{n(u_b,w_b,t+\tau)})},
\ee
\end{widetext}
where the overline means time averaging as $\overline{f(t)}=1/\Delta T \int_{-\Delta T/2}^{\Delta T/2}{f(t) dt}$.

\subsection{Assumptions}

Let us assume that there is a significantly large number of structures so that a statistical description is appropriate. For this description it is essential to know the distribution of the structure parameters ($u_0$, $w_0$, $t_0$). In our model we take these to be independent, uniform random variables, thus the probability density function is
\be
\label{uniform_dist}
  P(t_{0}) = \left\{
  \begin{array}{l l}
    \frac{1}{\Delta T} & \quad -\Delta T/2 \leq t_0 \leq \Delta T/2 \\
    0 & \quad \text{otherwise.}\\
  \end{array} \right.
\ee
A similar expression can be given for $u_0$ and $w_0$, but a physical meaning is still necessary, thus we attribute $\Delta T$ to the time length of the experimental signal and $\Delta U$, $\Delta W$ are the spatial extents of the observed poloidal plane. Due to the fact that the coherent structures vanish at much smaller than the size of the poloidal plane and the time length of the measurement, temporal and spatial averages can be taken as infinite integrals (e.g. $\int_{-\Delta T/2}^{\Delta T/2}{f(t) P(t) dt}\approx\int_{-\infty}^{\infty}{f(t) P(t) dt}$).

To simplify further calculations let us rewrite Eq. (\ref{cross_correlation_tot}) as
\begin{eqnarray}
\label{correlation_appr}
C_{a,b}(\tau)\equiv C(u_a,w_a,u_b,w_b,\tau)\\
\overline{(n_a(t)- \overline{n_a(t)}) (n_b(t+\tau)- \overline{n_b(t+\tau)})}= \nonumber\\
 \overline{n_a(t) n_b(t+\tau)}-\overline{n_a(t)}\cdot\overline{n_b(t+\tau)}.\nonumber
\end{eqnarray}
To reduce the complexity of future formulas let us also define the following quantity
\be
\label{kappa}
\kappa^2\equiv\frac{1}{\sigma_T^2}+\frac{v_u^2}{\sigma_U^2}+\frac{v_w^2}{\sigma_W^2},
\ee
which is the inverse of the characteristic decorrelation time \cite{Bencze_Zoletnik_PoP}. 

\subsection{Expected value of the total correlation function}

Using Eqs. (\ref{n_total}) and (\ref{correlation_appr}) the expected value of the cross correlation function (CCF) can be calculated, leading to the following expression
\begin{eqnarray}
\label{corr_anal_form}
\lang C_{a,b}(\tau)\rang=N \lang c_{a_i,b_i}(\tau)\rang+N(N-1)\lang c_{a_i,b_j}(\tau) \rang\\
-N \lang s_{a_i} s_{b_i} \rang-N(N-1)\lang s_{a_i}\rang\lang s_{b_j}\rang\nonumber,
\end{eqnarray}
where
\be
\label{sa}
s_{a_i}\equiv\overline{n_a(u_i,w_i,t)},
\ee
is the average contribution of the $i^{th}$ structure to the density in observation point $A$, and
\be
\label{cab}
c_{a_i,b_j}(\tau)\equiv\overline{n_a(u_i,w_i,t_i,t)n_b(u_j,w_j,t_j,t+\tau)},
\ee
 is the contribution to the CCF originating from two different coherent structures, called \textit{pair correlation function}. The individual terms of Eq. (\ref{corr_anal_form}) can be easily evaluated as they are basically Gaussian integrals. Thus
\be
\label{eval_sa}
s_{a_i}=\frac{\sqrt{2 \pi }}{\text{$\Delta T$} \kappa} e^{-\frac{\left(\text{$w_i$}-w_s(a,i)\right){}^2 \left(\text{$v_u$}^2 \text{$\sigma_T
   $}^2+\text{$\sigma_U$}^2\right)}{2 \kappa^2 \text{$\sigma_T$}^2 \text{$\sigma_U$}^2 \text{$\sigma_W$}^2}} e^{-\frac{\left(\text{$u_a$}-\text{$u_i$}\right){}^2}{2 \left(\text{$v_u$}^2
   \text{$\sigma_T$}^2+\text{$\sigma_U$}^2\right)}},
\ee
where
\be
\label{eq:ws_def}
w_s(a,i)\equiv\text{$w_a$}+\frac{\text{$v_u$} \text{$v_w$} \text{$\sigma_T$}^2 }{\text{$v_u$}^2 \text{$\sigma_T$}^2+\text{$\sigma_U$}^2}(\text{$u_i$}-\text{$u_a$}).
\ee

Meanwhile the pair correlation function for two structures is
\be
\label{cab_simpl}
c_{a_i,b_j}(\tau)=\frac{\sqrt{\pi }}{\kappa \text{$\Delta T$}}A_{i,j} f_{i,j}(\tau ),
\ee
where
\be
\label{Aij}
	A_{i,j}\equiv \frac{\text{$\Delta T$}^2 \kappa^2}{2 \pi } s_{a_i} s_{b_j},
\ee
%
\be
\label{fij}
f_{i,j}(\tau )\equiv e^{-\frac{1}{4} \kappa^2 \left(\tau -\tau_{i,j}\right)^2},
\ee
and
\begin{eqnarray}
\label{tauij}
\tau_{i,j}\equiv (t_j-t_i)+\frac{\text{$v_u$}}{ \kappa^2 \text{$\sigma_U$}^2} \left(u_i-u_j+\text{$u_b$}-\text{$u_a$}\right)\\
+\frac{\text{$v_w$}}{ \kappa^2\text{$\sigma_W$}^2}\left(w_i-w_j+\text{$w_b$}-\text{$w_a$}\right)\nonumber.
\end{eqnarray}

As the previous equations have shown, the results are rather complex, although still Gaussian. From now on only the most essential formulas will be presented to conserve space and allow the reader to follow the derivation. It follows from Eq. (\ref{corr_anal_form}) that the expected value of the CCF is
\begin{widetext}
\begin{eqnarray}
\label{corr_expval}
\lang C_{a,b}(\tau)\rang=N \frac{\pi ^{3/2} \text{$\sigma_T$} \text{$\sigma_U$} \text{$\sigma_W$}}{\text{$\Delta T$} \text{$\Delta U$} \text{$\Delta W$}} e^{-\frac{v^2 \text{$\sigma_T^2$} \sin^2{\beta}+\text{$\delta u$}^2 \text{$\sigma_W^2$}+\text{$\delta w$}^2 \text{$\sigma_U^2$}}{4 \kappa^2 \text{$\sigma _T^2$} \text{$\sigma_U^2$} \text{$\sigma_W^2$}}}
	\left[ e^{-\frac{1}{4} \kappa^2 \left(\tau-\lang\hat{D}\rang\right)^2}-\frac{2 \sqrt{\pi}}{\kappa \Delta T}\right],
\end{eqnarray}
\end{widetext}
where $\delta u\equiv u_b-u_a$, $\delta w\equiv w_b-w_a$, and $\kappa$ is set according to Eq. (\ref{kappa}), while $\lang\hat{D}\rang$ is the expected time delay -- the central quantity of the paper -- which will be defined in the next section in Eq. (\ref{tc_expval}). We also introduced $\beta$, which is the angle between the velocity vector ([$v_u$;$v_w$]) and the vector defined by the observation points ([$\text{$\delta u$}$;$\text{$\delta w$}$]), and $\text{$\delta l$}$ which is the distance between the observation points (see Fig. \ref{fig:model}).

Eq. (\ref{corr_expval}) also shows that $\kappa$ is in fact the characteristic time delay scale on which correlation vanishes, thus $\kappa$ is the decorrelation time.

\subsection{Time delay estimation and its variance}

In signal processing the position of the CCF peak -- from now on referred to as \textit{time delay estimate} (TDE) -- is essential in determining several key parameters of the turbulent structures (see Sec. \ref{sec:application}). The TDE (denoted as $\hat{D}$) can be derived by solving
\be
\label{TDE_deriv_eq}
\frac{d C_{a,b}}{d\tau}\bigg|_{\tau=\hat{D}}=\sum_{i,j}{\frac{d c_{a_i,b_j}}{d\tau}\bigg|_{\tau=\hat{D}}}=0.
\ee
Using Eq (\ref{corr_expval}) the expected TDE $\left(\lang\hat{D}\rang\right)$ becomes
\be
\label{tc_expval}
\lang \hat{D}\rang=\frac{\frac{v_u \delta u}{\sigma _U^2}+\frac{v_w \delta w}{\sigma _W^2}}{\kappa^2},
\ee
where $\kappa$ is defined according to Eq. (\ref{kappa}), $v_u=v_z \sin\alpha+v_r \cos\alpha$, $v_w=v_z \cos\alpha-v_r \sin\alpha$, $\delta u=\delta z \sin\alpha+\delta r \cos\alpha$ and $\delta w=\delta z \cos\alpha-\delta r \sin\alpha$ (see Fig. \ref{fig:model}).

It should be noted that a similar result was derived for the case of a single elliptical structure by Fedorczak et al. \cite{Fedorczak_tiltangle}, which can be considered the $\sigma_T\rightarrow\infty$ limit of this result. It can be shown that Eq. \ref{tc_expval} remains valid for not only Gaussian, but any other spatio-temporal distributions with elliptical contour surfaces.


The determination of time dependent parameters (e.g. flow velocity) based on TDE methods implies the usage of small time intervals for the calculation of cross-correlation function with reasonable time resolution. Thus a very valid question can be formulated: what are the relevant parameters determining the error (variance) of the calculation as a function of the time interval, or in other words for a given time resolution (frequency) and given error how long the time subintervals should be.

Based on the arguments in \cite{Tal} it is reasonable to assume that $\hat{D}$ is close to $\lang\hat{D}\rang$, which means that $c_{a_i,b_j}(\hat{D})$ can be approximated with a Taylor-series around $\lang\hat{D}\rang$. Taking a second order approximation of $c_{a_i,b_j}(\hat{D})$ and substituting it into Eq. (\ref{TDE_deriv_eq}) yields
\begin{eqnarray}
\sum_{i,j}{\frac{d c_{a_i,b_j}}{d\tau}\bigg|_{\tau=\lang\hat{D}\rang}}=\sum_{i,j}\frac{d c_{a_i,b_j}}{d\tau}\bigg|_{\tau=\lang\hat{D}\rang}+\\
\sum_{i,j}\frac{d^2 c_{a_i,b_j}}{d\tau^2}\bigg|_{\tau=\lang\hat{D}\rang}\left(\hat{D}-\lang\hat{D}\rang\right)=0 \nonumber.
\end{eqnarray}
Let us define the following quantity
\be
\label{delta_tauij}
\text{$\Delta \tau_{i,j}$}\equiv\tau_{i,j}-\lang\hat{D}\rang,
\ee
which is the difference between the position of the peak of the pair correlation function and the expected TDE. Using Eq. (\ref{delta_tauij}) we can derive $\hat{D}-\lang\hat{D}\rang$. Substituting the form of Eq. (\ref{cab_simpl}) and introducing $B_{i,j}\equiv f_{i,j}\left(\lang\hat{D}\rang\right)$ yields
\be
\label{TDE_diff}
\hat{D}-\lang\hat{D}\rang=\frac{\sum_{i,j}{A_{i,j}B_{i,j}\text{$\Delta \tau_{i,j}$}}}{\sum_{i,j}{A_{i,j}B_{i,j}\left(\frac{\text{$\Delta \tau_{i,j}$}}{\sigma^2}-1\right)}}.
\ee

Equation (\ref{TDE_diff}) is rather complex and our goal is to calculate its first and second moments, where we expect the first moment to be zero, while the second moment will be the variance of the TDE. Since $\hat{D}-\lang\hat{D}\rang$ is small, it is possible to expand the expression as
\be
\frac{Y}{X}=Y\left(\frac{1}{\lang X \rang}-\frac{1}{\lang X \rang^2}\left(X-\lang X \rang\right)+...\right),
\ee
where we denoted the denominator of Eq. (\ref{TDE_diff}) as $X$ and the numerator as $Y$. It can be shown that in the $N\rightarrow\infty$ limit
\be
\lang\left(X-\lang X \rang\right)^2\rang\sim \ord\left(\frac{1}{\kappa \Delta T}\right) \lang X \rang^2,
\ee
where $\frac{1}{\kappa \Delta T}=\frac{\frac{1}{\Delta T}}{\sqrt{\frac{1}{\sigma_T^2}+\frac{v_u^2}{\sigma_U^2}+\frac{v_w^2}{\sigma_W^2}}}\ll 1$. This means that a low order estimation around the expected value would be sufficient. It is of course just an intuitive argument as $Y$ and $X$ are not independent in this case (see Eq. (\ref{TDE_diff})). In the $N\rightarrow\infty$ limit the first moment of Eq. (\ref{TDE_diff}) gives zero in all orders of expansion as all terms are $\ord\left(\frac{1}{N}\right)$ or lower, while the second moment gives a finite value in zeroth order. It should be noted that taking only the zeroth order term is identical to assuming that the nominator ($Y$) and denominator ($X$) of Eq. (\ref{TDE_diff}) are independent. This gives the following expression for the TDE variance in the high density limit:
\be
\label{TDE_var_2}
\sigma_0^2\left(\hat{D}\right)=\frac{\sqrt{\frac{\pi }{2}}}{\kappa^{5} \text{$\Delta T$} \text{$\sigma_T$}^2}\left[\frac{\text{$\delta u$}^2}{\text{$\sigma_U$}^2}+\frac{\text{$\delta w$}^2}{\text{$\sigma_W$}^2}+\frac{\text{$\delta l$}^2\text{$\sigma_T$}^2 v^2}{\text{$\sigma_U$}^2 \text{$\sigma_W$}^2} \sin^2{\beta}\right],
\ee
where $\beta$ and $\text{$\delta l$}$ are the same as in Eq. (\ref{corr_expval}) (see Fig. \ref{fig:model}).

As in zeroth order the velocity is inversely proportional to the TDE, its error can be estimated as
\be
\sigma_v\approx v \frac{\lang \hat{D}\rang}{\sigma_0} \propto 1/\Delta T,
\ee
which clearly shows the trade-off between the accuracy of the velocity estimation and the frequency resolution.

The results in this section can be considered two-dimensional generalizations of the model presented in \cite{Tal}, but derived without further approximations. 

\section{Comparison with simulation}\label{sec:simulation}

To arrive at the result Eq. (\ref{TDE_var_2}) we have employed a number of approximations, thus, to test the correctness of our analytical predictions for the TDE variance, a numerical simulation code was developed in Matlab, which directly simulated the model depicted in Sec. \ref{sec:model}, then calculated the TDE and the orientation angle from the simulated signals. The simulation was rerun with a multitude of random initial conditions from which the statistics of the TDE and the tilt angle were derived. The structure parameters used by the simulation are detailed in Table \ref{tab:simparam}.

\begin{table}[htbp]
	\centering
		\begin{tabular}{ | c | c || c| c |}
		\hline
		$\text{$\Delta T$}$ & $2400\,\rm{\mu s}$ & $\text{$\sigma_T $}$ & $50\,\rm{\mu s}$  \\
		\hline
		$\text{$\Delta u$}$ & $200\,\rm{cm}$ & $\text{$\sigma_U $}$ & $2\,\rm{cm}$  \\
		\hline
		$\text{$\Delta w$}$ & $30\,\rm{cm}$ & $\text{$\sigma_W $}$ & $1\,\rm{cm}$  \\
		\hline
		$\text{$v_x$}$ & $0\,\rm{m/s}$ &  $\text{$\delta x$}$ & $1\,\rm{cm}$ \\
		\hline
		$\text{$v_y$}$ & $1000\,\rm{m/s}$ & $\text{$\delta y$}$ & $1\,\rm{cm}$  \\
		\hline
		$\alpha$ & $\pi/6$ &  &   \\
		\hline
		\end{tabular}
	\caption{Default parameters of numerical simulation.}
	\label{tab:simparam}
\end{table}

To study the transition into the high density limit it is useful to define the \textit{filling} quantity, which is the ratio of the volume occupied by coherent structures in parameter space to its total volume
\be
\label{filling}
\rm{filling}\equiv\left(8 N \sigma_T\sigma_U\sigma_W\right)/\left(\Delta U \Delta W \Delta T\right).
\ee
The filling value can be considered a time average of the so-called \textit{packing fraction}, which is the fraction of the poloidal surface occupied by coherent structures.

\begin{figure}[H]
\begin {center}
\includegraphics[width=\linewidth]{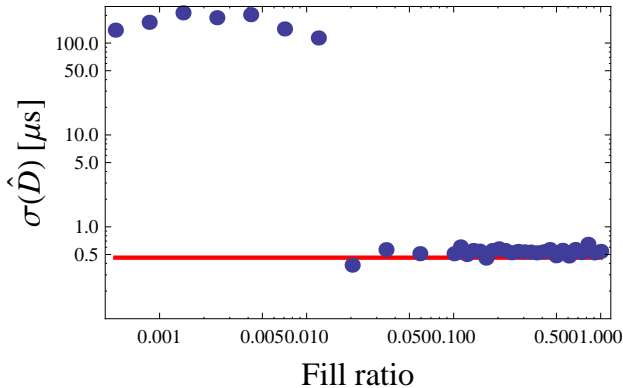}
\caption{Standard deviation of TDE using simulation results for different filling values (blue dots) compared to analytical prediction in the high density limit (Eq. (\ref{TDE_var_2}) -- red line). Filling is defined according to Eq. (\ref{filling}).}\label{fig:prev_compare_tau}
\end {center}
\end{figure}

Fig. \ref{fig:prev_compare_tau} shows the standard variation of the TDE as the filling increases. It is apparent from Fig. \ref{fig:prev_compare_tau} that the deviation of the TDE reaches the high density limit even for very low filling values ($\sim 2\%$), thus using the high density limit is justified in experimental situations, where the filling value is usually $\sim 10\%$ \cite{fillingexp}.

\begin{figure}[H]
\begin {center}
\includegraphics[width=\linewidth]{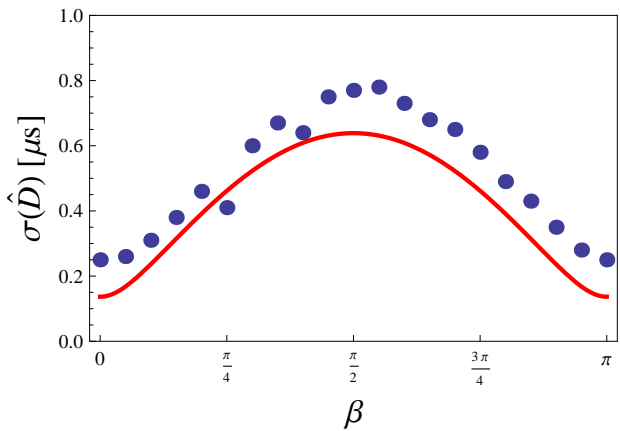}
\caption{Standard deviation of TDE in the high density limit as a function of $\beta$ (see Fig. \ref{fig:model}) according to simulation (blue dots) and analytical formula (red line).}\label{fig:betacompare}
\end {center}
\end{figure}

\begin{figure}[H]
\begin {center}
\includegraphics[width=\linewidth]{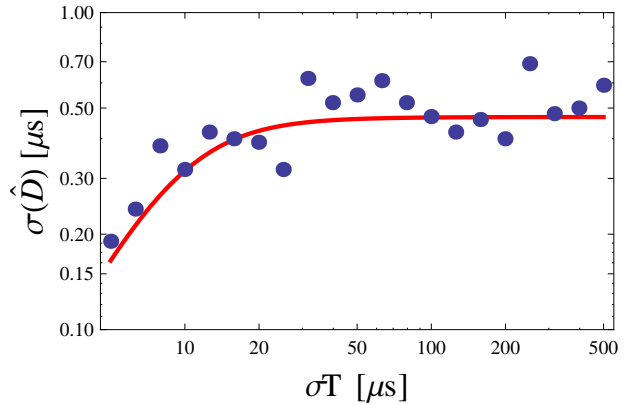}
\caption{ Standard deviation of TDE in the high density limit as a function of the structure's lifetime ($\sigma_T$) according to simulation (blue dots) and analytical formula (red line).}\label{fig:sigTcompare}
\end {center}
\end{figure}

\begin{figure}[H]
\begin {center}
\includegraphics[width=\linewidth]{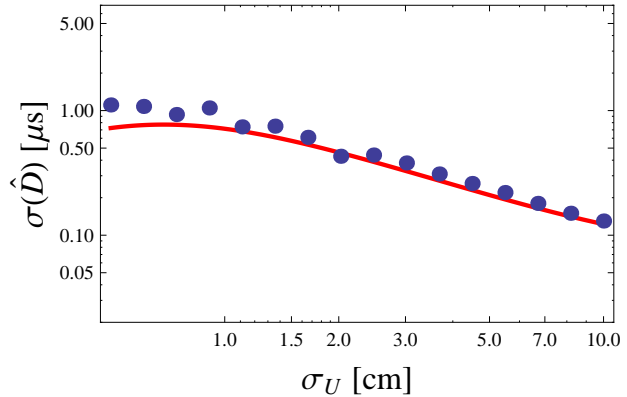}
\caption{ Standard deviation of TDE in the high density limit as a function of the structure's major axis ($\sigma_U$) according to simulation (blue dots) and analytical formula (red line).}\label{fig:sigUcompare}
\end {center}
\end{figure}

The parameter dependence of Eq. (\ref{TDE_var_2}) was also validated against simulations as shown on Figs. \ref{fig:betacompare}-\ref{fig:sigUcompare}.

It should be noted that the perturbations of the model discussed in Sec. \ref{sec:model} are non-physical in the sense that their spatial average is not zero, thus violating particle conservation. A better model would be, if the spatial distribution of coherent structures was not simply Gaussian, but polynomial times Gaussian. The simplest of such models is
\be
\label{eq:better_model}
n_i(u,w,t)=n_i^{\rm{Gauss}}(1-\hat{u}^2-\hat{w}^2),
\ee
where $n_i^{\rm{Gauss}}$ is the density perturbation from Eq. (\ref{blob_eq}), while $\hat{u}^2$ and $\hat{w}^2$ are the exponents of their respective Gaussians. Carrying out the analysis of Sec. \ref{sec:model} for this model would be challenging as the complexity of the previous formulas would drastically increase. Meanwhile numerical simulations showed that using this more accurate model causes no significant deviation from the TDE calculated in Sec. \ref{sec:model}.

\subsection{Standard deviation of orientation angle}

The TDE (Eq. (\ref{tc_expval})) of the correlation function depends on the tilt angle of structures ($\alpha$ on Fig. \ref{fig:model}) thus by measuring the TDE $\alpha$ -- among other parameters -- could be determined (see Sec. \ref{sec:application}). However, to ascertain the validity of those calculations knowing the variance of the calculated $\alpha$ is necessary. Using a linear estimation
\be
\label{alfa_diff_linear}
\Delta \alpha=\frac{d\alpha}{d \hat{D}}\sigma(\hat{D})+\ord\left(\sigma^2\left(\hat{D}\right)\right).
\ee
From Eq. (\ref{tc_expval}) the derivative can be easily calculated
\be
\label{alfa_deriv}
\frac{d\alpha}{d \hat{D}}=\left(\frac{d \hat{D}}{d\alpha}\right)^{-1}=\frac{\kappa^2 }{\left(\frac{1}{\sigma _U^2}-\frac{1}{\sigma _W^2}\right) \left(\delta w v_u+\delta u v_w-2 \hat{D} v_u v_w\right)}.
\ee
Using this result and Eq. (\ref{alfa_diff_linear}) the high density limit of the standard deviation of $\alpha$ yields
\begin{eqnarray}
\label{alfa_deviation_highdens}
	\sigma_0(\alpha)\approx \sqrt{\frac{\sqrt{\frac{\pi }{2}}}{\kappa \text{$\Delta T$} \text{$\sigma_T$}}}\times\\\frac{\sqrt{\frac{\text{$\delta u$}^2}{\text{$\sigma_U$}^2}+\frac{\text{$\delta w$}^2}{\text{$\sigma_W$}^2}+\frac{\text{$\delta l$}^2\text{$\sigma_T$}^2 v^2}{\text{$\sigma_U$}^2 \text{$\sigma_W$}^2} \sin^2{\beta}} }{\left(\frac{1}{\sigma _U^2}-\frac{1}{\sigma _W^2}\right) \left(\delta w v_u+\delta u v_w-2 \hat{D} v_u v_w\right)}\nonumber.
\end{eqnarray}

\begin{figure}[H]
\begin {center}
\includegraphics[width=\linewidth]{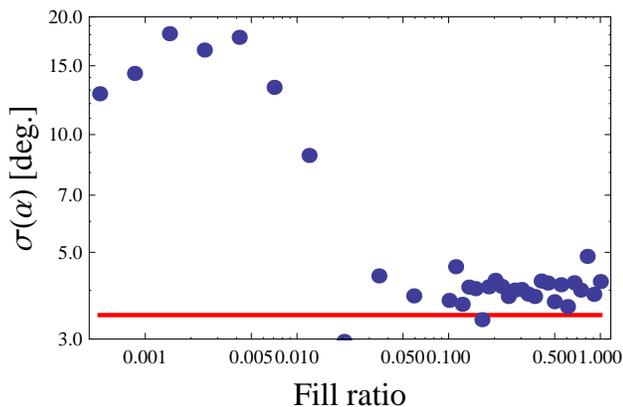}
\caption{Standard deviation of tilt angle values calculated from simulated signals for different filling values (blue dots) compared to the high density analytical prediction of Eq. (\ref{alfa_deviation_highdens}) (red line). Filling is defined according to Eq. (\ref{filling}). }\label{fig:simalfacompare}
\end {center}
\end{figure}

Fig. \ref{fig:simalfacompare} shows numerical results for different filling values along with the high density limit of $\sigma(\alpha)$ (Eq. (\ref{alfa_deviation_highdens})). Although the analytical formula of Eq. (\ref{alfa_deviation_highdens}) does not reproduce the simulation results perfectly -- due to the linear estimation used in Eq. (\ref{alfa_diff_linear}) -- it does give an order of magnitude estimate on the standard deviation of the angle.

\section{Application to TEXTOR data}\label{sec:application}

\begin{figure}[H]
\begin {center}
\includegraphics[width=\linewidth]{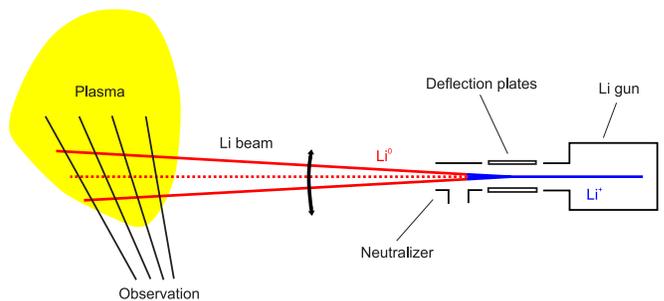}
\caption{Schematic of a quasi 2D measurement with Li BES.}\label{fig:quasi2d}
\end {center}
\end{figure}

The results from Sec. \ref{sec:model} allow a more detailed analysis of measured turbulence signals, for instance regarding the orientation of coherent structures. 
As a demonstration several parameters of turbulent structures in the TEXTOR tokamak ($R=1.75\,\rm{m}$; $a=0.47\,\rm{m}$; limited, circular plasma; $n_e=10^{19}\,\rm{m^{-3}}$) were calculated. For that purpose measured data from the Lithium Beam Emission Spectroscopy (Li-BES) \cite{Thomas_BES, Anda_EPS} diagnostic was used. In the examined discharge (\#113917, $I_p=350\,\rm{kA}$, $B_t=-1.9\,\rm{T}$) the diagnostic was in '\textit{fast deflection mode}', which means that during the discharge the beam was deflected by charged plates at high frequency before neutralization. This method allows the measurement of density fluctuations along not one but two beam lines hence it is called a '\textit{quasi-2D}' measurement \cite{quasi2d} (Fig. \ref{fig:quasi2d}).

After calculating the cross-correlation between individual channels, the time delay had to be determined as well. Unfortunately one of the disadvantages of the quasi-2D measurement is the greatly reduced time resolution ($2.4\,\rm{\mu s}$ in this case) which is of the same order of magnitude as the time delays ($\ord\left(3\,\rm{\mu s}\right)$). Thus the position of the peak was determined by fitting a parabola at the peak of the measured signal.

%

\subsection{Fitted results}

\begin{figure}[H]
\begin {center}
\includegraphics[width=\linewidth]{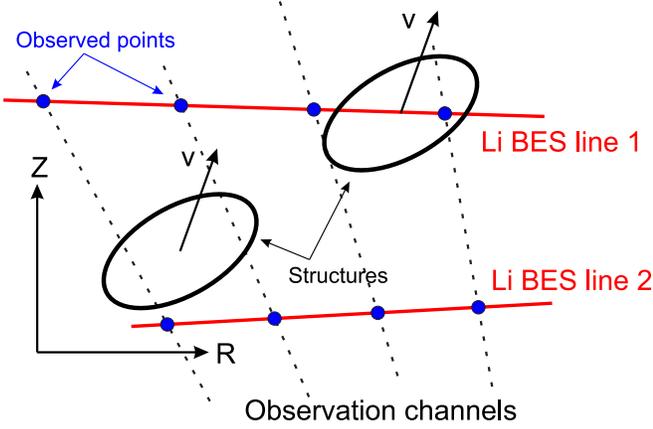}
\caption{Measurement configuration for TEXTOR quasi 2D Li BES.}\label{fig:meas_config}
\end {center}
\end{figure}

\begin{figure}[h]
\begin {center}
\includegraphics[width=\linewidth]{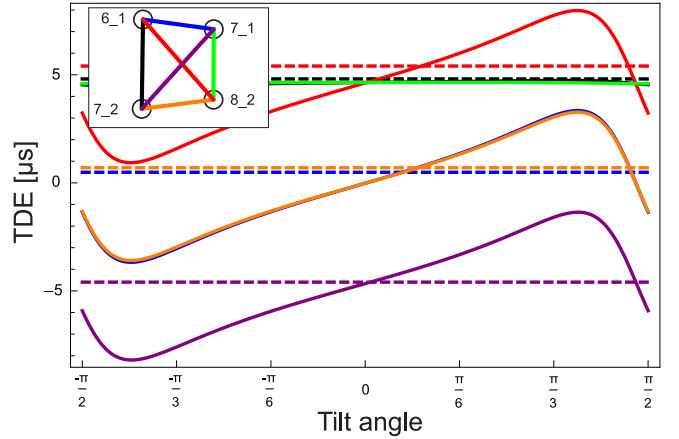}
\caption{Dependence of the TDE on the tilt angle in a realistic scenario. The dashed lines show measured TDEs for \#113917 around BES channel 6, while the solid lines show TDE curves according to Eq. (\ref{tc_expval}). The rest of the parameters are taken from the results of the fitting procedure mentioned before. Due to the measurement geometry, only 4 equations are independent, of which only 3 are have significant angle dependence.}\label{fig:TDE_4points}
\end {center}
\end{figure}

In the model we adopted the turbulent structures have 6 independent parameters ($\alpha$, $v_r$, $v_z$, $\sigma_T$, $\sigma_U$, $\sigma_W$). It is known that the turbulent structures have a poloidal velocity of several km/s-s while the poloidal distance between observation points are several cm-s, which implies a characteristic time of flight of $10\,\rm{\mu s}$, which is much shorter than the lifetime of the structures (thus we can take the $\sigma_T\rightarrow\infty$ limit). This simplify the expected TDE of formula of Eq. (\ref{tc_expval}) to
\be
\label{tc_expval_fit}
\lang \hat{D}\rang\approx\frac{v_u \delta u+v_w \delta w \epsilon^2}{v_u^2+v_w^2\epsilon^2},
\ee
where $\epsilon=\sigma _U/\sigma _W$ is the elongation of the structure. This means that only 4 parameters need to be fitted ($\alpha$, $v_z$, $v_r$, $\epsilon$). To be able to fit these parameters the TDE of cross correlations between 4 neighboring points (Fig. \ref{fig:meas_config}) were calculated (6 equations). As only the position differences of the observation points matter in Eq. (\ref{tc_expval}), the geometry of the quasi-2D measurement is rather problematic due to the parallel lines, which reduce the number of independent equations to 4 (see \ref{fig:TDE_4points}). 

One could consider taking into account the CCFs between far away points, but that is generally not feasible as the signal-to-noise ratio would be too small for non-neighboring points, while the parameters (e.g. velocity) are not necessarily constant on larger scales ($~ 2\,\rm{cm}$). Combined with the non-linear relation between parameters and the TDE in Eq. (\ref{tc_expval}), fitting the TDEs by itself can not provide a unique solution for all parameters, but it can restrict their possible values. According to our numerical tests -- with exact TDEs-- taking the measured decorrelation time ($\kappa$ in Eq. (\ref{corr_expval})) into account leads to unique solutions.

The parameters are fitted numerically using an iterative method (standard Levenberg-Marquardt algorithm) from randomly chosen initial parameters. The convergence is established using $\chi_{\rm{red}}^2<1$, where $\chi_{\rm{red}}$ is the reduced $\chi^2$. The errors are calculated from the statistical error of the TDE and the systematic error of the calibration of distances between observed points, which -- in this case -- are much more significant.

The fitting procedure also takes advantage of the fact that coherent structures in the plasma edge primarily propagate In the poloidal direction. In case of \#113917 the zeroth order approximation of their poloidal velocity is $v_z\approx \Delta z/\hat{D}\approx 3.5\,\rm{km/s}$, while the apparent radial velocity is $v_r^{\rm{app}}=\Delta r/\hat{D}\approx 10\,\rm{km/s}$. This means that the high apparent radial velocity can only be explained by the presence of a tilt, which is responsible for the major part of $v_r^{\rm{app}}$. This is fortunate, because in general the effects of radial propagation and structure tilt are hard to distinguish, but in this case the effects of $v_r$ are negligible.

\begin{figure}[h]
\begin {center}
\includegraphics[width=\linewidth]{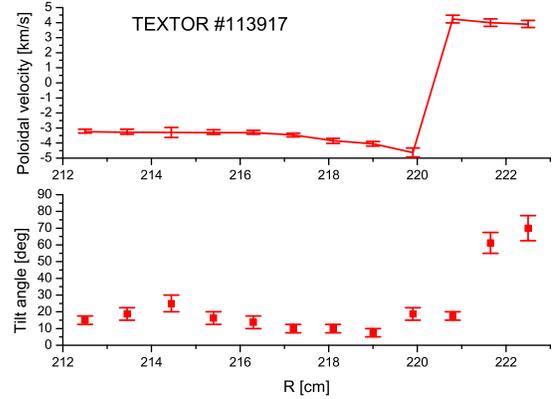}
\caption{Fitted poloidal velocities and tilt angles for TEXTOR discharge \#113917. The error bars are determined by $\chi_{\rm{red}}^2\leq 1$.}\label{fig:Fit_result}
\end {center}
\end{figure}

Figure \ref{fig:Fit_result} shows that the tilt angle of coherent structures is around 10-20 degrees, while Fig. \ref{fig:TDE_4points} shows there are other solutions around 90 degrees. The reason these were discarded is, that the TDE changes around this value are very sharply (see Fig. \ref{fig:TDE_4points}), which means that virtually no scatter in the orientation of structures could be allowed in order to reproduce the measured TDEs. 

The fitting also determined that $\epsilon\in [1.9; 2.9]$, which means that the structures were significantly elongated.

It is important to note that at $R>220\,\rm{cm}$ the velocity gradient steepens drastically, causing a significant deformation of the structures, thus violating the assumption of spatially constant structure parameters between observed points, thus fitted parameters in that range are likely erroneous. 

The fitting results were compared against the results from the TEXTOR Correlation Reflectometry (CR) \cite{Kramer_TEXTOR}. The CR results show, that poloidal velocity at $R=216\,\rm{cm}$ is $-3.2\,\rm{km/s}$, while the tilt angle is 5.1 degrees. Although there is a discrepancy between this angle and Fig. \ref{fig:Fit_result}, it is explained by the fact, that BES and CR measurements are carried out at different poloidal positions.


\section{Conclusion}

Time Delay Estimation (TDE) is one of most commonly used method to study turbulent structures in fusion plasmas. To describe the coherent structures at the plasma edge a simple two-dimensional Gaussian model is considered, which can be seen as the generalization of the model of Tal et al. \cite{Tal}. The key statistical quantities of the model were calculated and it was established that in the high density limit the variance of the cross correlation function's (CCF) peak -- the time delay estimate (TDE) -- is low, while its dependence on structure parameters is relatively simple, making it a good candidate to determine the parameters of coherent structures. A possible application of the model was demonstrated on a TEXTOR discharge, where the radial profiles of several key blob-parameters (poloidal velocity, tilt angle, elongation) were determined. A systematic application of the method will be detailed in a follow-up publication.

\end{document}